# Model-independent Limits from Spin-dependent WIMP Dark Matter Experiments

F. Giuliani[*] and TA Girard

*Centro de Física Nuclear, Universidade de Lisboa, 1649-003 Lisboa, Portugal*

Spin-dependent WIMP searches have traditionally presented results within an odd group approximation and by suppressing one of the spin-dependent interaction cross sections. We here elaborate on a model-independent analysis in which spin-dependent interactions with both protons and neutrons are simultaneously considered. Within this approach, equivalent current limits on the WIMP-nucleon interaction at WIMP mass of 50 GeV/c$^2$ are either $\sigma_p \leq 0.7$ pb, $\sigma_n \leq 0.2$ pb or $|a_p| \leq 0.4$, $|a_n| \leq 0.7$ depending on the choice of cross section or coupling strength representation. These limits become less restrictive for either larger or smaller masses; they are less restrictive than those from the traditional odd group approximation regardless of WIMP mass. Combination of experimental results are seen to produce significantly more restrictive limits than those obtained from any single experiment. Experiments traditionally considered spin-independent are moreover found to severely limit the spin-dependent phase space. The extension of this analysis to the case of positive signal experiments is explored.

## I. INTRODUCTION

The missing matter of the Universe continues to be among the forefront interests of current physics activity. A halo of non-relativistic, weakly interacting massive particles (WIMPs) trapped in a galactic gravitational well is one of the natural candidates for this missing matter. Numerous searches for evidence of WIMP existence have been made, either via direct observation of nuclear recoils generated by their elastic scattering off target nuclei, or indirectly via observation of WIMP annihilation signatures.

Direct searches have been traditionally characterized by whether the scattering is spin-independent (in which case the coupling scales with the mass of the target nucleus) or spin-dependent (in which case the coupling is nonzero only if the spin of the nucleus is nonvanishing). For spin-dependent interactions, the traditional analysis is to consider only a WIMP-proton or WIMP-neutron coupling, whether by assuming the WIMP to interact with only the odd group of detector nucleons (leaving the even group as inert spectators) or by simply setting the other coupling equal to zero. In either case, an experiment using only odd Z isotopes cannot constrain a theory predicting a WIMP-neutron coupling, *i.e.* the exclusion results quoted for the experiment are WIMP model dependent. Nuclear structure calculations however show that the even group of nucleons might have a non-negligible (though subdominant) spin. Moreover, theory does not restrict the WIMP-proton and WIMP-neutron coupling strengths, especially for the neutralino since this is a superposition of higgsinos and gauginos.

The impact of a model-independent analysis [1], in which spin-dependent interactions with both protons and neutrons are simultaneously allowed, was recently considered by one of us in a Letter [2]. The analysis has examine [5] whether simultaneous spin-dependent couplings might reconcile the DAMA/NaI annual modulation signal with its exclusion by DAMA/Xe-2 [6].

As stated in Ref. [2], the description can be formulated in either a coupling strength or cross section representation. Here, we elaborate on each, which was not possible in the Letter because of space restrictions. Although entirely equivalent, each yields seemingly different exclusion plots on the WIMP phase space. Generally, allowing both WIMP-neutron and WIMP-proton couplings weakens the restrictions obtained by assuming one or the other equals zero, but allows more restrictive limits by combining experimental results.

After a brief review of the usual odd group framework [7], the model-independent framework is described in Sections IIA and IIB. Section III presents the results of various spin-dependent WIMP searches in the model-independent framework, with the DAMA/NaI results described separately in Section IV which extends the model-independent framework to the case of experiments with a positive WIMP signal. Section V contains a discussion of the considerations, together with criteria to assist in the definition of new search experiment detectors.

## II. THE BASIC FRAMEWORK

The general form of an effective non-differential lowest order lagrangian for the elastic scattering of a generic spin 1/2 WIMP on nucleons is [8]:

$$\mathcal{L} = 4\sqrt{2}G_F \left[ \chi^+ \vec{\sigma}\chi (a_p p^+ \vec{\sigma} p + a_n n^+ \vec{\sigma} n) + \right.$$

franck@cii.fc.ul.pt

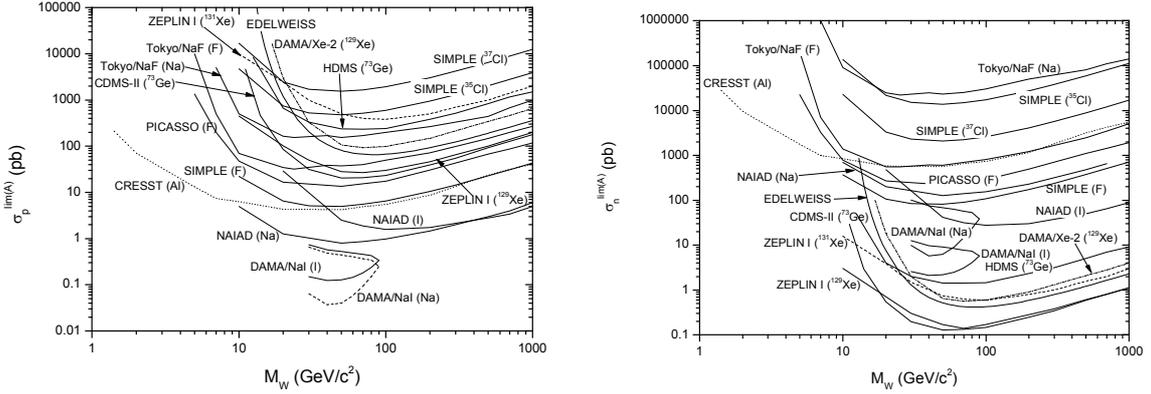

FIG.1: $\sigma_p^{lim(A)}$ (left) and $\sigma_n^{lim(A)}$ (right) vs $M_W$ for various elements, from the experiments of Table I. All curves are calculated at 90% C.L., except for DAMA/NaI which is a 3σ result.

$$\chi^+\chi(f_p p^+ p + f_n n^+ n)\,]\qquad(1)$$

where p, n and χ are the proton, neutron and WIMP two-component Weyl spinors respectively, $\bar{\sigma}$ are the Pauli matrices, $G_F$ is the Fermi coupling constant, and $a_{p,n}$ are the effective proton (neutron) coupling strengths (coupling constants in units of $\sqrt{2}G_F$) for the spin-dependent interaction sector, while $f_{p,n}$ are the spin-independent coupling strengths.

A detailed discussion, starting from a relativistic lagrangian, of the validity of Eq. (1) can be found in Ref. [8]. Here, we simply point out that Eq. (1) can be derived on very general theoretical grounds without assuming supersymmetry [8], by starting from the lowest order approximation for the interaction lagrangian as justified by the interaction weakness.

On the basis of Eq. (1) and of the nuclear shell model, the general zero momentum transfer WIMP-nuclide spin-dependent cross section $\sigma_A$ for elastic scattering on a nucleus of mass number A is then given at tree level by [7-9]:

$$\sigma_A = \frac{32}{\pi}G_F^2\mu_A^2\big(a_p\langle S_p\rangle + a_n\langle S_n\rangle\big)^2\frac{J+1}{J}\ ,\qquad(2)$$

where $\langle S_{p,n}\rangle$ are the expectation values of the proton (neutron) group's spin, $\mu_A$ is the WIMP-nuclide reduced mass, and $J$ is the total nuclear spin.

After converting the experimental data to limits on $\sigma_A$, experiments usually quote limits only for a single nucleon interaction by calculating, for each sensitive nuclide, one of the following cross sections:

$$\sigma_{p,n}^{lim(A)} = \frac{3}{4}\frac{J}{J+1}\frac{\mu_{p,n}^2}{\mu_A^2}\frac{\sigma_A^{lim}}{\langle S_{p,n}\rangle^2}\qquad(3)$$

where $\sigma_A^{lim}$ is the upper limit on $\sigma_A$ obtained by experimental data, $\mu_{p,n}$ are the WIMP-proton (WIMP-neutron) reduced mass, and $\sigma_{p,n}^{lim(A)}$ are the proton and neutron cross section limits when $a_{n,p}<S_{n,p}>=0$ respectively. The $\sigma_A^{lim}$ is evaluated by attributing the entire counting rate to the nuclide A, so that the $\sigma_{p,n}^{lim(A)}$ are overestimated by a factor $\frac{1}{f_A}$, where $f_A$ is the atomic fraction of the isotope of mass number A. Since $\sum_A f_A = 1$, and $f_A \propto \frac{1}{\sigma_A^{lim(A)}}$, the constant of proportionality is taken as a more refined WIMP-proton (WIMP-neutron) cross section limit $\sigma_{p,n}^{lim}$, obtained from limits in Eq. (3) by using the relation [7] $\frac{1}{\sigma_{p,n}^{lim}} = \sum_A \frac{1}{\sigma_n^{lim(A)}}$.

TABLE I: currently active dark matter search experiments.

| experiment | material | exposure (kgd) | mass (kg) | Ref. |
|---|---|---|---|---|
| NAIAD | NaI | 3879 | 46 | [10] |
| DAMA/NaI | NaI | 57986 | 100 | [11] |
| DAMA/Xe-2 | $^{129}$Xe | 1763 | 6.5 | [12] |
| ZEPLIN-I | Xe | 293 | 3.2 | [13] |
| EDELWEISS | Ge | 62 | 0.96 | [14] |
| CDMS-II | Ge | 19 | 1 | [15] |
| CRESST-I | $Al_2O_3$ | 1.51 | 0.262 | [16] |
| SIMPLE | $C_2Cl F_5$ | 0.6 | 0.06 | [4] |
| PICASSO | $C_3F_8,C_4F_{10}$ | 0.056 | 0.00134 | [17] |
| Tokyo/NaF | NaF | 3.38 | 0.176 | [3] |
| CRESST-II | $CaWO_4$ | 10.7 | 0.3 | [18] |
| HDMS | $^{73}$Ge (86%) | 85.5 | 0.202 | [19] |

Plots of $\sigma_{p,n}^{lim(A)}$ vs WIMP mass ($M_W$) for the experiments listed in Table I are shown in Fig. 1. This figure cannot be directly compared with the traditional odd group exclusion plots, which are of $\sigma_{p,n}^{lim}$, except in the case of single isotope experiments, for which the two plots coincide. The curves for SIMPLE, NAIAD, ZEPLIN-I and the Tokyo collaboration are taken from Refs. [3,4,10,13], while those for the remaining experiments are calculated on the basis of Refs. [13,12,14-17].

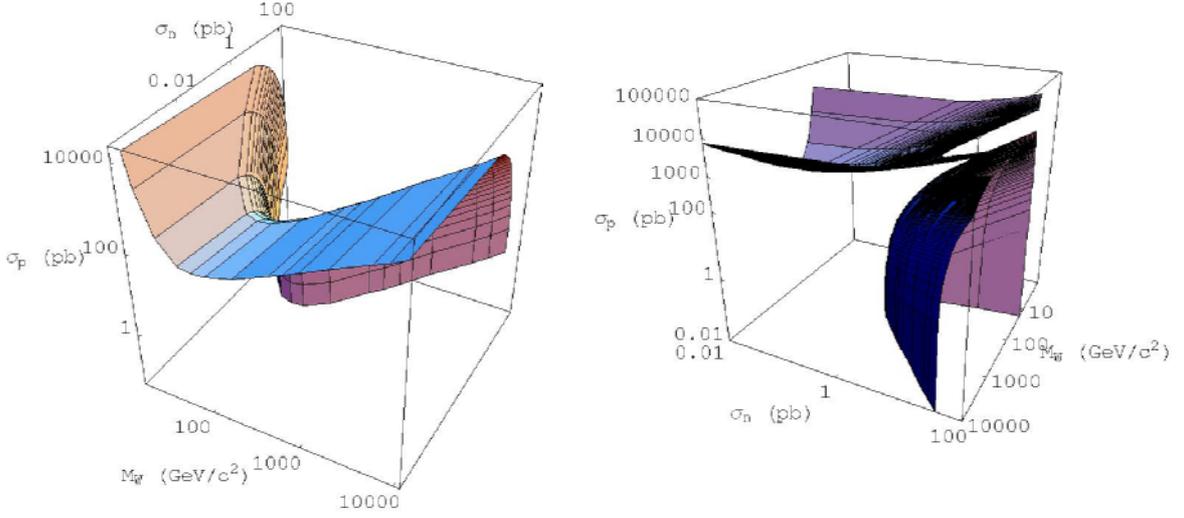

FIG. 2: 3D exclusion plots in the cross section representation, for DAMA/Xe-2 with $\frac{a_n <S_n>}{a_p <S_p>} > 0$ (left) and $\frac{a_n <S_n>}{a_p <S_p>} < 0$ (right).

The calculation of $\sigma_{p,n}^{lim(A)}$ of Na and I for DAMA/NaI required recalculation of $\sigma_A^{lim}$ for Na and I from a mixed model [11] $\sigma_p$ and $\sigma_n$, followed by re-application of Eq. (3). This is not completely rigorous, but a good approximation. The DAMA/Xe-2 experiment, instead, uses almost pure $^{129}$Xe, so it is sufficient to multiply the $\sigma_n^{lim(A)}$ reported in Ref. 12 by $\left(\frac{\mu_p <S_n>}{\mu_n <S_p>}\right)^2$ to get also $\sigma_p^{lim(A)}$. CRESST-I (Al$_2$O$_3$) did not report a $\sigma_n^{lim(A)}$, but since $^{16}$O is an even-even, spinless and doubly magic nucleus with no magnetic moment [20], its spin-dependent response has been neglected, and the exclusion of Ref. 16 has been assumed to result from Al only. Similar arguments apply to PICASSO, in which only F is sensitive to spin-dependent WIMPs, and to both CDMS-II [15] and EDELWEISS [14], whose active elements contain only one spin-dependent isotope.

Traditionally, experiments based on natural Si and Ge such as CDMS-II and EDELWEISS would not be included in the present discussion, because almost all of their isotopes are spinless even-even nuclei, hence essentially spin-dependent insensitive. However $^{73}$Ge (7.8% of natural Ge) and $^{29}$Si (4.67% of natural Si) do contribute in spite of their small abundance. Also included are the recently reported results from the HDMS experiment enriched to 86% in $^{73}$Ge [19]. As for CRESST-II (CaWO4), its spin-dependent isotopes are $^{43}$Ca (0.135% of natural Ca), $^{17}$O (0.038% of natural O), and $^{183}$W (14.3% of natural W), which *together with* the small $<S_{p,n}>$ of $^{183}$W (Table II) makes the experiment an essentially spin-independent WIMP search.

As stated in Ref. [2], model-independent limits on spin-dependent WIMP-nucleon interaction can be formulated either in terms of nucleon cross sections (cross section representation) or the coupling strengths $a_{p,n}$. We consider each of these equivalent formulations, first in the simpler case of a single spin-dependent sensitive isotope and then in the general case of a compound active material.

### A. Cross section representation

#### 1. Single isotope experiments

Eq. (2) for the single proton (neutron) cross section $\sigma_p$ ($\sigma_n$) becomes:

$$\sigma_{p,n} = \frac{32}{\pi} G_F^2 \mu_{p,n}^2 \frac{3}{4} a_{p,n}^2 \quad . \tag{4}$$

Substituting Eqs. (2)-(4) into the obvious relation $\frac{\sigma_A}{\sigma_A^{lim}} \leq 1$, it can be shown that [1]:

$$\left(\sqrt{\frac{\sigma_p}{\sigma_p^{lim(A)}}} \pm \sqrt{\frac{\sigma_n}{\sigma_n^{lim(A)}}}\right)^2 \leq 1 \quad , \tag{5}$$

where the sign of the addition in parenthesis is that of the $\frac{a_n <S_n>}{a_p <S_p>}$ ratio. Eq. (5) indicates the need for a separate treatment of the cases $\frac{a_n}{a_p} > 0$ and $\frac{a_n}{a_p} < 0$.

Since $\sigma_{p,n}^{lim(A)}$ are functions of $M_W$ only, Eq. (5) describes a relation in a three dimensional phase space, as shown in Fig. 2. There are two allowed volumes, depending on the sign of $\frac{a_n <S_n>}{a_p <S_p>}$, both delimited by the indicated sheets and by the coordinate planes. As indicated in Fig. 2, mass cuts near the minima of Fig. 1 (~ 50 GeV/c$^2$, which is in the DAMA/NaI preferred range [11] generally provide the smallest $\sigma_{p,n}$; cuts above or below this mass imply larger allowed areas.

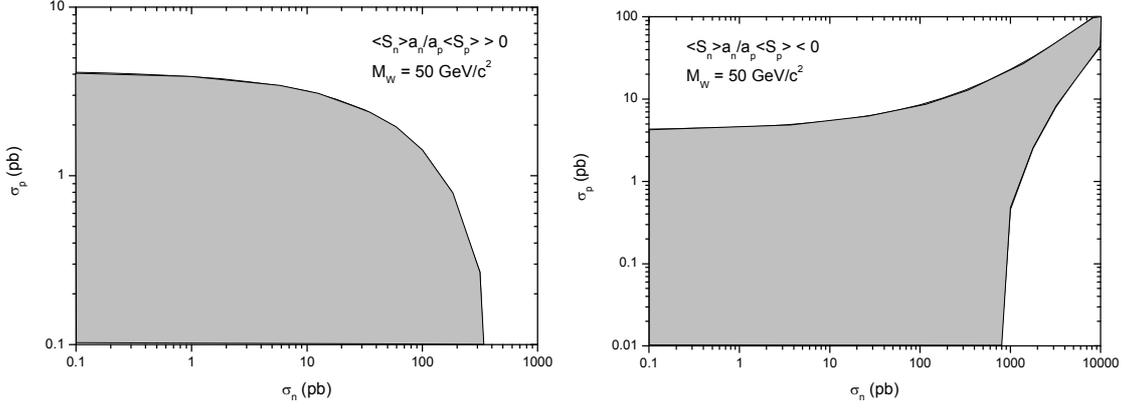

FIG. 3: exclusion plots in the $\sigma_p$-$\sigma_n$ plane at 50 GeV/c$^2$ WIMP mass, for CRESST-I and $\frac{a_n <S_n>}{a_p <S_p>} > 0$ (left) and $\frac{a_n <S_n>}{a_p <S_p>} < 0$ (right). Since $\frac{<S_n>}{<S_p>} > 0$, the left figure is for $\frac{a_n}{a_p} > 0$, the right for $\frac{a_n}{a_p} < 0$. The allowed areas are shaded.

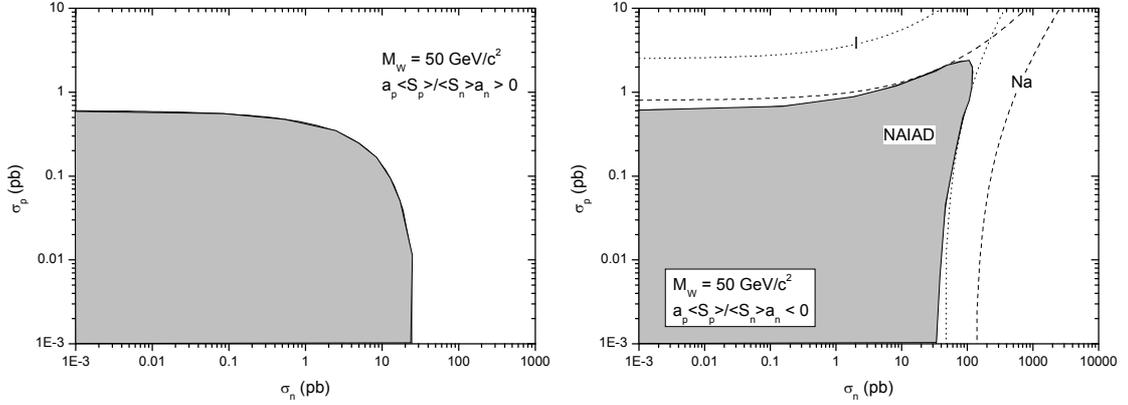

FIG. 4: typical allowed regions (shaded) in the $\sigma_p$ - $\sigma_n$ plane at 50 GeV/c$^2$ WIMP mass, for NAIAD with $\frac{a_n <S_n>}{a_p <S_p>} > 0$ (left) and $\frac{a_n <S_n>}{a_p <S_p>} < 0$ (right).

A more useful form is to project Fig. 2 onto the $\sigma_p$-$\sigma_n$ plane for a given $M_W$ [1-4]. In this framework, Eq. (5) constrains the allowed values of $\sigma_p$ and $\sigma_n$ to a closed or an open region in the projected phase space, again depending on the sign of $\frac{a_n <S_n>}{a_p <S_p>}$. The typical shape of these regions is shown in Fig. 3, taking as an example the CRESST-I results for $M_W$=50 GeV/c$^2$.

While for $\frac{a_n <S_n>}{a_p <S_p>} > 0$ the allowed region (shaded) is finite and limited by the $\sigma_p$ and $\sigma_n$ axes and by a rounded curve, for $\frac{a_n <S_n>}{a_p <S_p>} < 0$ the allowed region is unbounded for large $\sigma_{p,n}$. This is due to the sign in Eq. (5) being negative, so that $\sigma_p$ can grow indefinitely provided that the value of $\sigma_n$ keeps close enough to that of $\sigma_p$, which quenches $\sigma_A$ for theoretical scenarios such that $\frac{a_n}{a_p} = -\frac{<S_p>}{<S_n>}$. Consequently, limits from a single isotope experiment cannot constrain these scenarios at tree level. When $\frac{a_n}{a_p} = -\frac{<S_p>}{<S_n>}$, the tree level nucleus cross section is suppressed, which makes second order (or higher) terms possibly non-negligible [21].

### 2. Multi-nuclide experiments

For experiments with more than one nuclide in the active material, $\sigma_A$ is evaluated by attributing the entire counting rate to the nuclide A, so that the $\sigma_{p,n}^{lim(A)}$ are overestimated by a factor $\frac{1}{f_A}$. Consequently the limits of Eq. (5) $\sigma_p$ and $\sigma_n$ become, for each nuclide,

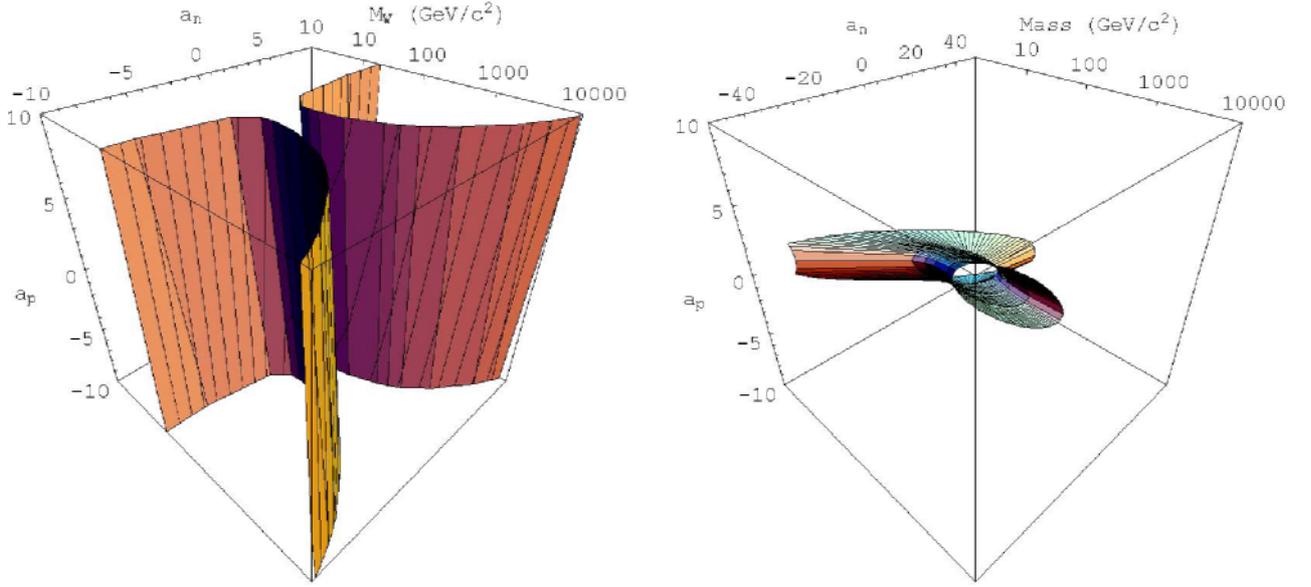

FIG. 5: 3D exclusion plots in the coupling strength representation for (left) DAMA/Xe-2 and (right) NAIAD.

$$\left( \sqrt{\frac{\sigma_p}{\sigma_p^{lim(A)}}} \pm \sqrt{\frac{\sigma_n}{\sigma_n^{lim(A)}}} \right)^2 \leq f_A \quad . \qquad (6)$$

Since $\sum_A f_A = 1$, then

$$\sum_A \left( \sqrt{\frac{\sigma_p}{\sigma_p^{lim(A)}}} \pm \sqrt{\frac{\sigma_n}{\sigma_n^{lim(A)}}} \right)^2 \leq 1 \quad , \qquad (7)$$

where the sum is intended over the nuclear species present in the detector's sensitive volume.

Eq. (7) makes the protuberance of Fig. 3 finite, even if a "poor" nuclide like Na or Cl (*i.e.* one only weakly sensitive to the WIMP-neutron interaction) is included, as demonstrated in Fig. 4 using NAIAD's result at $M_W$=50 GeV/$c^2$, along with the limits it would have with only Na or I. These are shown by the shaded, dashed and dotted contours of Fig. 4, respectively. The equivalence of Eqs. (6) and (7) is due to $\sigma_p$ and $\sigma_n$ being the same in all squares summed in Eq. (7).

Each point satisfying Eq. (7) must also belong to the dashed and dotted regions, hence the overall allowed region is contained in their intersection. As long as this intersection is finite, the overall allowed region must be closed, and multi-nuclide experiments are able to set model-independent constraints on theory.

This also holds when more than two nuclei are involved, and when there are two with opposite sign of $\frac{\langle S_n \rangle}{\langle S_p \rangle}$, so that at least one of the intersecting regions is finite.

### B. Coupling strength representation

The distinction between the cases $a_n/a_p > 0$ and $a_n/a_p < 0$ can be removed by rewriting Eqs. (5)-(7) as

$$\sum_A \left( \frac{a_p}{\sqrt{\sigma_p^{lim(A)}}} \pm \frac{a_n}{\sqrt{\sigma_n^{lim(A)}}} \right)^2 \leq \frac{\pi}{24 G_F^2 \mu_p^2} \quad , \qquad (8)$$

where the small difference between $m_p$ and $m_n$ is neglected such that $\sigma_{p,n} \sim \frac{24}{\pi} G_F^2 \mu_p^2 a_{p,n}^2$. The sign of the addition in parenthesis is now that of the $\frac{\langle S_n \rangle}{\langle S_p \rangle}$ ratio, since $a_{p,n}$ retain their signs.

The parameter space remains 3-dimensional, with the allowed regions now delimited by two sheets symmetric with respect to the WIMP mass axis for single nuclide experiments, and by (closed surfaces shaped as) tubes coaxial with the $M_W$ axis for multi-nuclide experiments, as seen in Fig. 5.

Again, a useful representation is provided by projecting onto the $a_p$ - $a_n$ plane for a given $M_W$. In this case however, Eq. (8) can be rewritten as

$$\alpha a_p^2 + 2\beta a_p a_n + \gamma a_n^2 \leq \frac{\pi}{24 G_F^2 \mu_p^2} \quad , \qquad (9)$$

which represents a conic in the $a_p$ -$a_n$ plane, with coefficients $\alpha$, $\beta$ and $\gamma$ given by:

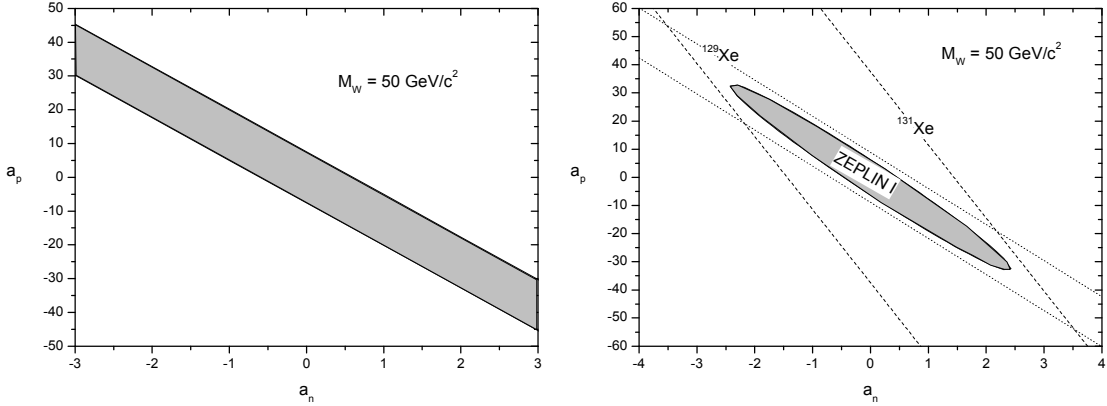

FIG. 6: 2D exclusion plots in the coupling strength representation for (left) CDMS-II and (right) ZEPLIN-I. The dashed and dotted lines result from attributing the entire rate to a single isotope (see text).

$$
\begin{cases}
\alpha = \sum_A \dfrac{1}{\sigma_p^{lim(A)}} \\[2mm]
\beta = \sum_A \pm \dfrac{1}{\sqrt{\sigma_p^{lim(A)} \sigma_n^{lim(A)}}} \\[2mm]
\gamma = \sum_A \dfrac{1}{\sigma_n^{lim(A)}}
\end{cases} . \qquad (10)
$$

The sign of each term in the summation for $\beta$ is determined by the sign of the $\frac{<S_n>}{<S_p>}$ ratio for the isotope of mass number A.

The coefficients $\alpha$, $\beta$ and $\gamma$ determine the contour of the results in Fig. 6, since the major and minor semiaxis $b_M$ and $b_m$ are proportional to the inverse eigenvalues of the matrix M=$\begin{pmatrix} \alpha & \beta \\ \beta & \gamma \end{pmatrix}$ obtained from the coefficients of the second degree terms of Eq. (9). The permitted region becomes the inside of a conic, which in the case of a single nuclide degenerates to a band between two parallel straight lines (Fig. 6), whose stiffness is $-\frac{<S_n>}{<S_p>}$. This translates in the coupling strength representation to the inability of an experiment using a single sensitive isotope to fully constrain all tree level theoretical scenarios.

For multi-nuclide experiments, the conic of Eq. (8) is an ellipse. In fact, Eq. (8) implies *a fortiori* $\left( \frac{a_p}{\sqrt{\sigma_p^{lim(A)}}} \pm \frac{a_n}{\sqrt{\sigma_n^{lim(A)}}} \right)^2 \leq \frac{\pi}{24 G_F^2 \mu_p^2}$ which is a single isotope band for each target nuclide. The overall conic must lie inside the intersection of all these bands. As a consequence, in the zero momentum transfer limit, the region allowed by any experiment is never a parabola nor a hyperbola. The argument of Section II A 2, that allowed regions of multi-nuclide experiments are limited, is easily translated within the coupling strength representation to the more restrictive condition

$$
\left( \frac{a_p}{\sqrt{\sigma_p^{lim(A)}}} \pm \frac{a_n}{\sqrt{\sigma_n^{lim(A)}}} \right)^2 \leq \frac{\pi}{24 G_F^2 \mu_p^2} f_A \qquad (11)
$$

for each nuclide. The corresponding bands in the $a_p$ - $a_n$ plane are shown in Fig. 6 for the case of the ZEPLIN-I result at $M_W = 50$ GeV/c$^2$, as dashed ($^{129}$Xe) and dotted ($^{131}$Xe). The intersection of these bands contains the overall allowed region. Since the various bands are generally not parallel, their intersection is usually finite. Hence the overall allowed region (shaded) must be a finite conic, as seen in Fig. 6.

## III. LIMITS FROM COMBINED EXPERIMENTS

Fig. 7 shows the shaded regions in the cross section representation allowed by each of the experiments considered here (except DAMA/NaI, treated separately in Section IV because of its positive signal) for $M_W = 20$, 50 and 100 GeV/c$^2$ in the two cases of $\frac{a_p}{a_n} < 0$ (left) and $\frac{a_p}{a_n} > 0$ (right). Note that for each sign of $\frac{a_p}{a_n}$, $\frac{a_p <S_p>}{a_n <S_n>}$ can have either sign.

In all cases, the overall allowed area is determined by intersecting NAIAD and either ZEPLIN-I or CDMS-II. The infinite protuberance of single nuclide experiments has no influence on their intersection with other experiments, which is always a limited area.

Note that CDMS-II, with only 19 kgd exposure and 7.7% of its bolometer active mass spin-dependent sensitive, is at the same level of the 293 kgd of the ZEPLIN-I scintillator, in spite of the larger fraction (47.6%) of spin-dependent sensitive active mass. This difference in exposure to reach the same level is reproduced by the difference between EDELWEISS 62 kgd with Ge bolometers and the DAMA/Xe-2 1763 kgd with a liquid Xe scintillator, and can be attributed to the different detection techniques. The difference in performance between ZEPLIN-I and DAMA/Xe-2 may be due to improvements in discrimination techniques, like the use of veto detectors.

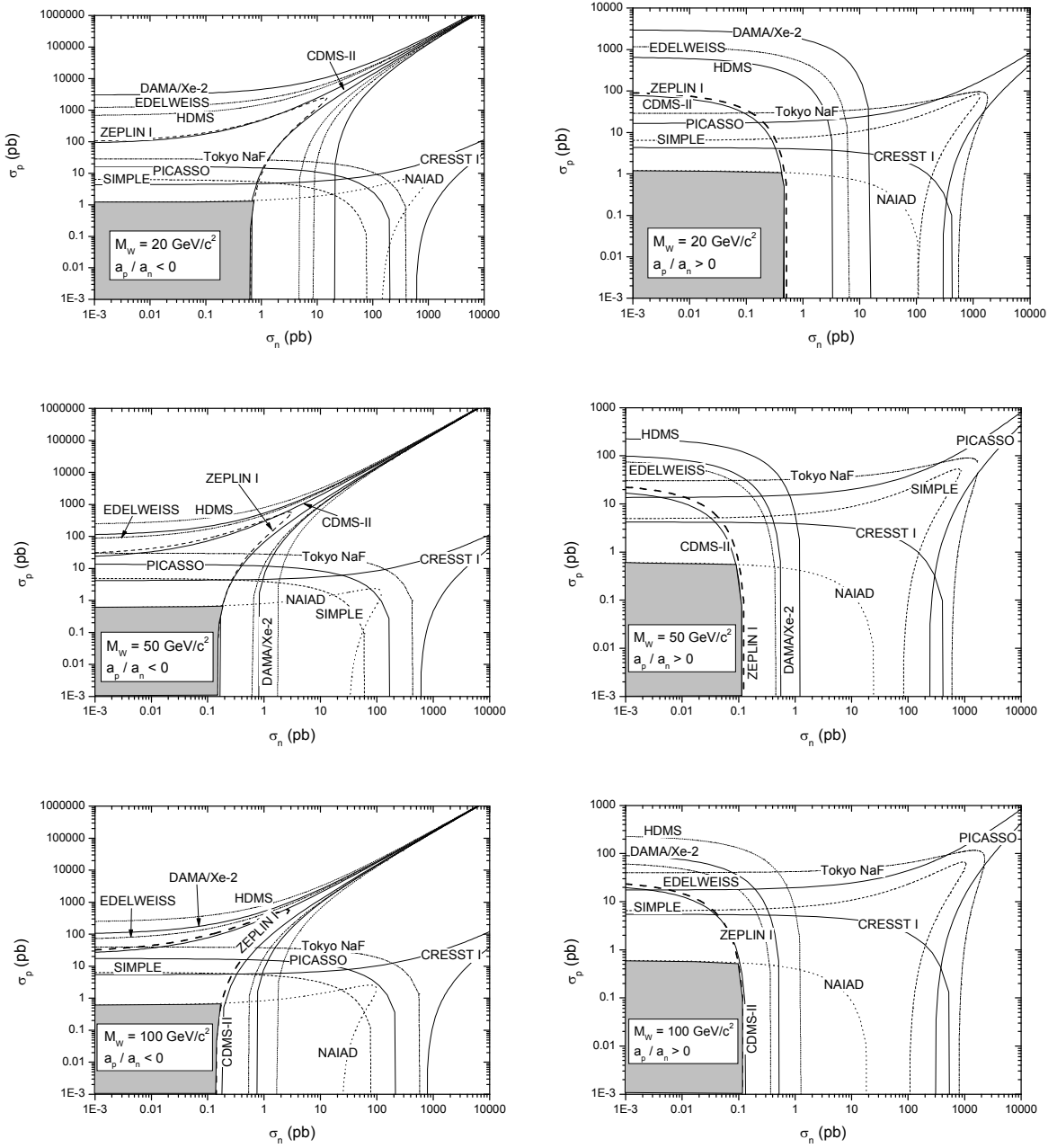

FIG. 7: Allowed regions (shaded) in the $\sigma_p$-$\sigma_n$ plane for $M_W = 20$, 50 and 100 GeV/c² with $\frac{a_p}{a_n} < 0$ (left) and $\frac{a_p}{a_n} > 0$ (right).

To the contrary, the fact that ZEPLIN-I provides at $M_W = 100$ GeV/c² limits more stringent than CDMS-II is a simple consequence of Eq. (2): the cross section of an isotope with mass number A is proportional to the square of the WIMP-isotope reduced mass $\mu_A$, which is maximum when $M_W$ matches the isotope mass. So, at $M_W$ = 100 GeV/c², ¹²⁹Xe and ¹³¹Xe have a cross section larger than ⁷³Ge.

The overall limits are, at $M_W = 50$ GeV/c², $\sigma_n \leq 0.2$ pb and $\sigma_p \leq 0.7$ pb. We however recall from Fig. 2 that for both low and high $M_W$ the limits become significantly larger. This is understood both on the basis of Eq. (2), because the WIMP-nuclide reduced mass is maximum when $M_W$ matches the nuclide's mass, and at low $M_W$ because the recoil energies tend to fall below the detector sensitivity threshold.

The advantage of the coupling strength representation is shown in Fig. 8 for the same values of $M_W$. Because of the difference in sign between the $\frac{<S_n>}{<S_p>}$ ratios of F and I, the angle between the ellipses of F-based and I-based experiments is large, allowing for an intersection significantly smaller than the original ellipses of each experiment. The 293 kgd result of ZEPLIN-I only provides a band delimited by two parallel lines, almost vertical due to the low $<S_p>$, so that the result cuts a very thin slice of the NAIAD ellipse. The slight mismatch between CDMS-II and ZEPLIN-I is due to the larger abundance of spin-dependent isotopes in Xe (47.6% vs 7.7% of Ge). In contrast, the 61.95 kgd EDELWEISS has a performance slightly better than the 1763 kgd DAMA/Xe-2, in spite of its 100% $^{129}$Xe. DAMA/Xe-2 did not use veto, which seems to give ZEPLIN-I a sensitivity improvement able to compensate for the factor 200 in exposure.

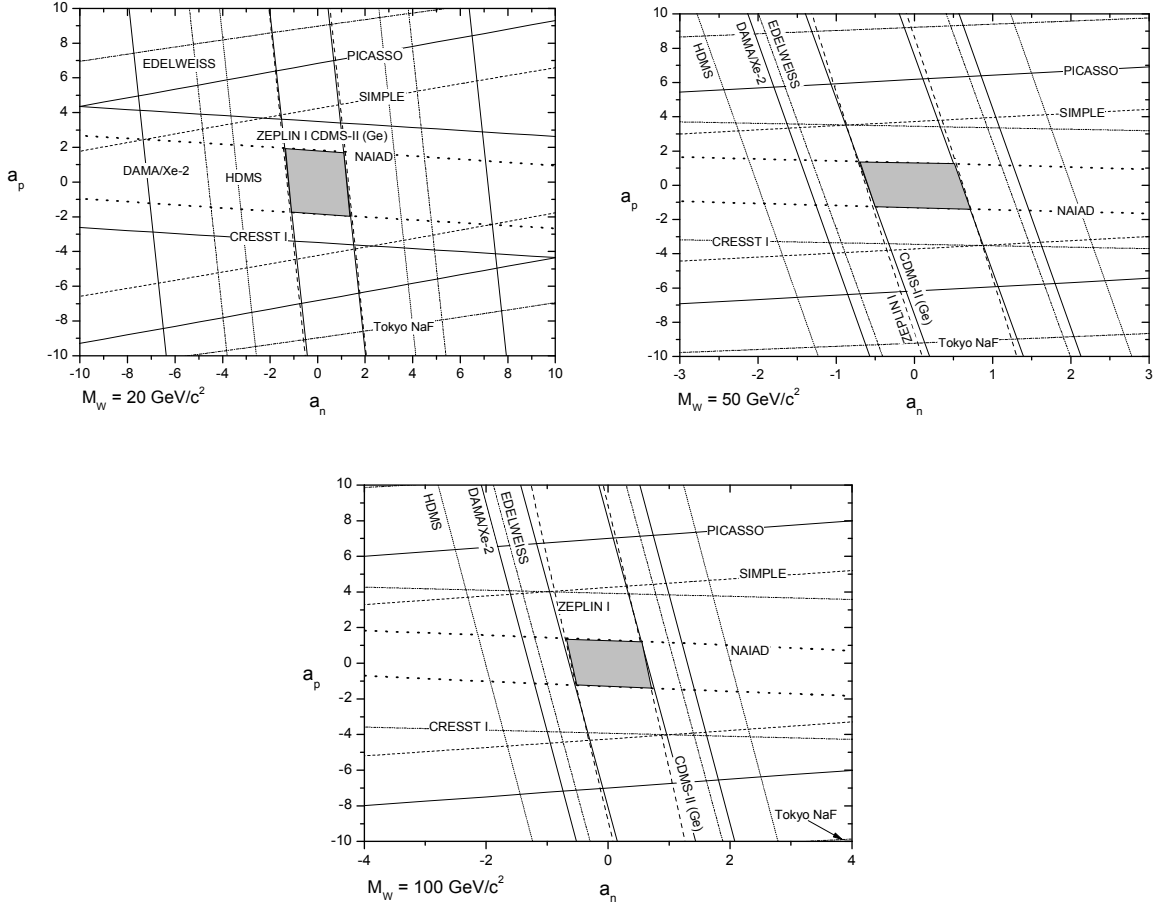

FIG. 8: Exclusion plots for $M_W$ = 20, 50 and 100 GeV/c$^2$ in the coupling strength representation. The overall allowed region, given by the intersection of all experiments, is shaded.

The resulting limits at $M_W$ = 50 GeV/c$^2$, $|a_n| \leq 0.7$ and $|a_p| \leq 1.4$, which via Eq. (4) are equivalent to the $\sigma_{p,n}$ from Fig. 7, do not yet exclude a WIMP whose spin-dependent interaction strength equals that of the ordinary weak interaction. Note that Eq. (2) has been written for Majorana WIMPs, like neutralinos, so if the true WIMP were a Dirac spinor, the ordinary weak interaction strength would correspond to half the values of $a_{p,n}$.

## IV. EXPERIMENTS WITH A POSITIVE SIGNAL

If an experiment has a positive WIMP signal, both an upper ($\sigma_A^{lim}$) and a lower ($\sigma_A^{lim\,inf}$) limit on $\sigma_A$ can be set, so that:

$$\frac{\sigma_A}{\sigma_A^{lim}} \leq f_A \leq \frac{\sigma_A}{\sigma_A^{lim\,inf}} \quad . \tag{12}$$

where the isotopic fractions $f_A$ are again due to the overestimate of attributing the entire counting rate to atomic specie A only, which affects both $\sigma_A^{lim}$ and $\sigma_A^{lim\,inf}$ by a factor $\frac{l}{f_A}$. Analogy with Eq. (3) suggests the introduction of the auxiliary cross sections:

$$\sigma_{p,n}^{lim\,inf(A)} = \frac{3}{4}\frac{J}{J+1}\frac{\mu_{p,n}^2}{\mu_A^2}\frac{\sigma_A^{lim\,inf}}{<S_{p,n}>} \quad , \tag{13}$$

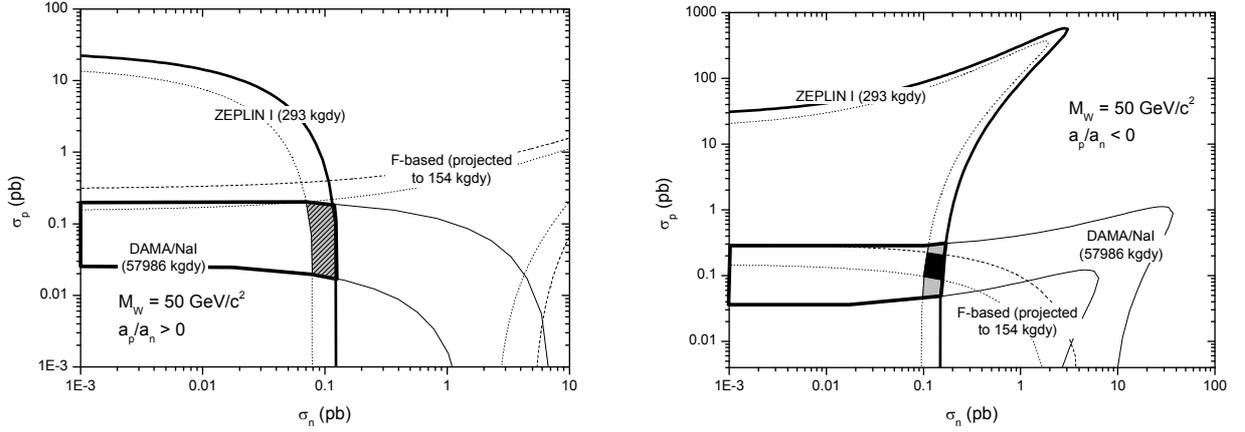

FIG. 9: Exclusion plots for positive signal experiments at $M_W = 50$ GeV/c² in the cross section representation. The thick border region is the intersection of DAMA/NaI's 3σ exclusion plot with the current ZEPLIN-I 90% C.L. result. The hatched areas are obtained by attributing an arbitrary positive signal to ZEPLIN-I, while the F-based experiment is a similar projection from SIMPLE. The overall allowed region, given by the intersection of all experiments, is solid, and in this example only exists for negative $\frac{a_p}{a_n}$.

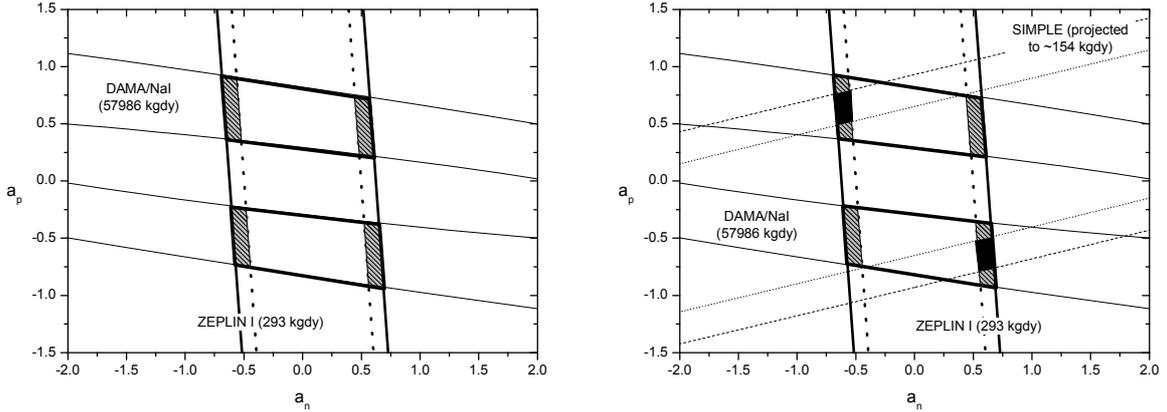

FIG. 10: Exclusion plots for positive signal experiments at $M_W = 50$ GeV/c² in the coupling strength representation. The thick border regions are the intersection of DAMA/NaI's 3σ exclusion plot with the current ZEPLIN-I 90% C.L. result. The hatched areas are obtained by attributing an arbitrary positive signal to ZEPLIN-I, while the F-based experiment is a similar projection from SIMPLE. The overall allowed region, given by the intersection of all experiments, is solid.

which allows completion of Eqs. (7, 8) with:

$$
\begin{cases}
\sum_A \left( \sqrt{\dfrac{\sigma_p}{\sigma_A^{lim\,inf}}} \pm \sqrt{\dfrac{\sigma_n}{\sigma_A^{lim\,inf}}} \right)^2 \geq 1 \\[3mm]
\sum_A \left( \dfrac{a_p}{\sigma_A^{lim\,inf}} \pm \dfrac{a_n}{\sigma_A^{lim\,inf}} \right)^2 \geq \dfrac{\pi}{24 G_F^2 \mu_p^2}
\end{cases} \quad (14)
$$

The first of these extra equations transforms the cross section representation of a multi-nuclide experiment into a curved band, whose extremities are on the $\sigma_p$ and $\sigma_n$ axes, as evident from Fig. 9, where only multi-nuclide experiments are displayed. The exclusion plot from a single nuclide experiment remains unbounded, as it becomes a pair of curved bands which do not intersect each other. One of these bands starts from the $\sigma_p$ axis and is delimited by two concave, nonintersecting curves, while the other starts from the $\sigma_n$ axis and is delimited by two convex, nonintersecting curves.

The second of Eqs. (14) transforms the ellipses or bands in the $a_p$-$a_n$ plane into elliptical "shells" and pairs of bands symmetric with respect to the origin (Fig. 10). Of course, WIMP masses such that the lower limits of Eq. (14) are incompatible with the upper limits (Eqs. (7), (8)) are excluded. Figs. 9, 10 show the regions allowed by DAMA/NaI and ZEPLIN-I for $M_W = 50$ GeV/c² in both representations. The intersection of the two results yields two closed regions, shown as thick borders. Should also ZEPLIN-I detect a WIMP signal, its allowed regions in the cross section representation will also become shells, whose inner contour is shown in Figs. 9 as two dotted lines. Clearly, there will be two distinct intersection areas, one for $\frac{a_p}{a_n} > 0$, the other for $\frac{a_p}{a_n} < 0$.

In the coupling strength representation of Fig. 10, the ZEPLIN-I allowed shell is almost perpendicular to DAMA/NaI's, yielding four allowed areas symmetric with respect to the origin of the $a_p$ - $a_n$ plane. These correspond to the two areas of the cross section representation. This symmetry and doubling are due to the fact that $a_{p,n} \propto \pm\sqrt{\sigma_{p,n}}$ , so that each point in the coupling strength representation corresponds to two symmetric points in the cross section representation.

Since in Figs. 9 and 10 the external contours of ZEPLIN-I are the actual current limits, the limits from DAMA/NaI and ZEPLIN-I for $M_W = 50$ GeV become $0 \leq |a_n| \leq 0.7$, $0.2 \leq |a_p| \leq 0.9$, or $0 \leq \sigma_n \leq 0.17$, $0.02 \leq \sigma_p \leq 0.31$ pb. Note that the WIMP-nucleon coupling strengths are still compatible with $|a_{p,n}| = 1$, i.e. an ordinary weak interaction strength.

To reduce the number of allowed regions, a third experiment with a different orientation is needed, such as the F-based experiments of SIMPLE and PICASSO. Figs. 9, 10 also contain a projection for these (dashed), along with an hypothetical positive signal (dotted). In this case, the F-based experiments would remove two of the hatched areas, allowing only the two solid regions symmetric with respect to the origin. The two regions correspond to positive (repulsive) and negative (attractive) WIMP-nucleon interaction energy.

## V.    DISCUSSION

As evident from Section III, the most restrictive limits on spin-dependent WIMP existence arise from the intersection of the results of multiple experiments, de facto reducing the number of "key" experiments capable of having significant impact on the issue.

To explore the impact of the model-independent framework on experiment design considerations, it is useful to define a "relative sensitivity" parameter D:

$$D = \begin{cases} \left(\dfrac{m_n}{m_p}\right)^2 \left(\dfrac{<S_p>}{<S_n>}\right)^2 \sim \dfrac{\sigma_n}{\sigma_p} & if \ \sigma_n < \sigma_p \\[4mm] \left(\dfrac{m_p}{mn}\right)^2 \left(\dfrac{<S_n>}{<S_p>}\right)^2 \sim \dfrac{\sigma_p}{\sigma_n} & if \ \sigma_n \geq \sigma_p \end{cases} , \quad (15)$$

When $M_W >> m_{n,p}$, D measures the ratio of "sensitivity" of the same nucleus to WIMP-proton and WIMP-neutron interaction. If $D \sim 0$, the model-independent treatment generally adds only a small correction to the odd group approximation and the band in Fig. 6 tends to be parallel to the $a_p$ ($a_n$) axis; if $D \sim 1$, the element is as sensitive to spin-dependent WIMP-proton as to WIMP-neutron interactions, so the band's stiffness is $\sim$

$\pm$ 1. Hence D is a measure of how heavily the refined model-independent analysis corrects more simplistic ones. Fig. 11 shows the values of D for a variety of elements commonly used in direct spin-dependent WIMP searches. Isotopes below the dashed line have $\frac{<S_n>}{<S_p>} < 0.1$ or $\frac{<S_n>}{<S_p>} > 10$, i.e. produce in Fig. 8 quasi-horizontal or quasi-vertical bands. High values of D, like the 0.74 of $^{37}$Cl, tend to produce $\sim \pm \pi/4$ orientations.

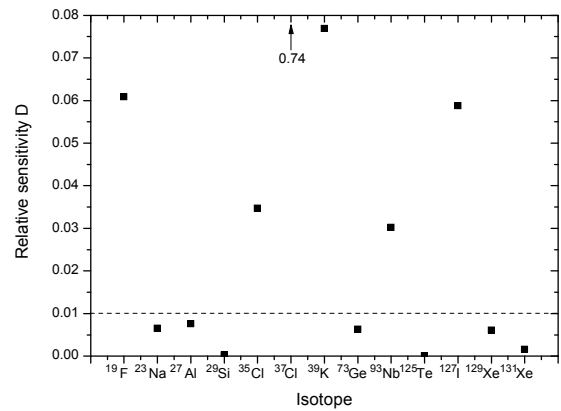

FIG. 11: "Relative sensitivity" D for various M nuclides. In the coupling strength representation, D is the modulus of the stiffness of the region permitted by an experiment using only that isotope. Isotopes below the dashed line yield a quasi-horizontal or -vertical permitted region.

The spread in sensitivity of Fig. 11 confirms the absence of any correlation between D and Z. However, it is worth to recall that $<S_n>$ and $<S_p>$ are not measured, but calculated by fitting nuclear models to known nuclear data, so the values of D are also nuclear model dependent. An example of such nuclear model dependence is provided by $^{35}$Cl [9], for which $<S_p>$= -0.15 in the odd group approximation, but only -0.059 in a refined shell model. This is due to the fact that most of the Cl $\frac{3}{2}$ total nuclear spin is contributed by the nucleons' orbital angular momenta. Since such phenomenon also occurs with $^{39}$K [22], it should reasonably occur with $^{37}$Cl, which has the same neutron number as $^{39}$K. $^{35}$Cl, $^{37}$Cl and $^{39}$K are "nearby" isotopes with low gyromagnetic ratio. The fact that the single nucleon spin gyromagnetic ratios are higher than the corresponding orbital gyromagnetic ratios indicates high orbital angular momenta, both for the proton and the neutron group. This is particularly striking for $^{39}$K, that possesses a non-negligible $<S_n>$ of 0.05 in spite of having a magic number of neutrons (closed neutron shell). Spin and angular momentum values can be found in the literature for $^{35}$Cl and $^{39}$K, but not for $^{37}$Cl. Since both $^{37}$Cl and $^{39}$K have 20 neutrons and similar number of protons, it is reasonable to assume that the neutron group of both isotopes is in the same state, hence that the neutron group spin and angular momentum of $^{37}$Cl are the same as calculated for $^{39}$K [22]. This allows calculation of the proton group spin and angular momentum from the measured values of its nuclear spin (3/2) and magnetic moment [20]. In fact:

$$\mu = g_p <S_p> + <L_p> + g_n<S_n> \qquad (16)$$
$$J = <S_p> + <L_p> + <S_n> + <L_n>$$

where $\mu$ is the magnetic moment in nuclear magnetons, the $<L_p>$ and $<L_n>$ are the expectation values of the proton and neutron group orbital angular momentum, $g_p$ and $g_n$ are the (vacuum) proton and neutron's gyromagnetic ratios, and the proton's orbital angular momentum gyromagnetic ratio has been taken equal to 1, as in vacuum. If $<L_n>$ and $<S_n>$ are known, it is easy to show that

$$< S_p > = \frac{\mu - J + (1 - g_n) < S_n > + < L_n >}{g_p - 1} \qquad . \quad (17)$$

Among the nuclides considered in Table II, those providing limits most affected by use of the model-independent scheme are $^{39}$K, $^{19}$F, $^{127}$I and $^{37}$Cl. For $^{39}$K, $^{19}$F and $^{127}$I, at zero momentum transfer, the limits on $\sigma_n$ differ by a factor 10-20 from those on $\sigma_p$, while for elements below the dashed line of Fig. 11 the difference is over two orders of magnitude. For $^{37}$Cl, the difference is a factor of $\sim 2$.

High relative sensitivity does not mean that an isotope is highly sensitive, e.g. $^{39}$K and $^{37}$Cl have a good relative sensitivity but both $<S_p>$ and $<S_n>$ are low. This and $^{40}$K radioactivity make K a poor choice for a sensitive WIMP detector. On the other hand, $^{93}$Nb has both spin values better than $^{127}$I (Table II), but its relative neutron sensitivity is lower.

The problem of evaluating a single experiment's constraint capabilities is addressed by the conic sections of the coupling strength representation, since the coefficients $\alpha$, $\beta$ and $\gamma$ determine the shape of the ellipse in the phase space. It is clear that an experiment designed to have equal sensitivity to all WIMP candidates produces circular permitted regions in the $a_p$ - $a_n$ plane, i.e. zero eccentricity ($\varepsilon$) ellipses. Since $\varepsilon \equiv \sqrt{1 - \left(\frac{b_m}{b_M}\right)^2}$ ,

$$\varepsilon = \sqrt{\frac{2\sqrt{(\alpha - \gamma)^2 + 4\beta^2}}{\alpha + \beta + \sqrt{(\alpha - \gamma)^2 + 4\beta^2}}} \qquad . \qquad (18)$$

The condition $\varepsilon = 0$ is then equivalent $\alpha = \gamma$ and $\beta = 0$, which translated in terms of $\sigma_{p,n}^{lim(A)}$ means balancing the detector composition by mixing odd Z and odd A-Z nuclei to get $\alpha = \gamma$, and by using nuclei with different signs of $\frac{<S_n>}{<S_p>}$ in order to get cancellations in the sum for $\beta$. Notice that $\beta = 0$ corresponds to an ellipse with axis parallel to the coordinate axis. Unfortunately, the $\sigma_{p,n}^{lim(A)}$ depend on the WIMP mass in a nonlinear fashion, so a specific detector composition may yield $\alpha = \gamma$ and $\beta = 0$ for some possible $M_W$ values, but not for all.

TABLE II: relevant spin values for the nuclides in this paper.

| nucleus | Z | J | $<S_p>$ | $<S_n>$ | D | Ref. |
|---|---|---|---|---|---|---|
| $^{17}$O [a] | 8 | 5/2 | 0 | 0.495 | 0 | |
| $^{19}$F | 9 | 1/2 | 0.441 | -0.109 | 0.0609 | [1] |
| $^{23}$Na | 11 | 3/2 | 0.248 | 0.020 | 0.0065 | [23] |
| $^{27}$Al | 13 | 5/2 | 0.343 | 0.030 | 0.0076 | [22] |
| $^{29}$Si | 14 | 1/2 | -0.002 | 0.130 | 0.0002 | [1] |
| $^{35}$Cl | 17 | 3/2 | -0.059 | -0.011 [b] | 0.0347 | [9] |
| $^{37}$Cl [a] | 17 | 3/2 | -0.178 | 0 | 0 | [4] |
| $^{37}$Cl [c] | 17 | 3/2 | -0.058 | 0.050 | 0.7411 | [4] |
| $^{39}$K | 19 | 3/2 | -0.18 | 0.050 | 0.0769 | [22] |
| $^{73}$Ge | 32 | 9/2 | 0.030 | 0.378 | 0.0063 | [1] |
| $^{93}$Nb | 41 | 9/2 | 0.460 | 0.080 | 0.0302 | [1] |
| $^{123}$Te | 52 | 1/2 | 0.001 | 0.287 | 1e-05 | [23] |
| $^{127}$I | 53 | 5/2 | 0.309 | 0.075 | 0.0588 | [23] |
| $^{129}$Xe | 54 | 1/2 | 0.028 | 0.359 | 0.0061 | [1] |
| $^{131}$Xe | 54 | 3/2 | -0.009 | -0.227 | 0.0016 | [23] |
| $^{183}$W [a] | 74 | 1/2 | 0 | -0.031 | 0 | |

[a] odd group approximation, calculated using data from Ref. [20].
[b] In Ref. [2] this value was mistyped.
[c] Evaluated by assuming the same neutron group spin and angular momentum as $^{39}$K (see text).

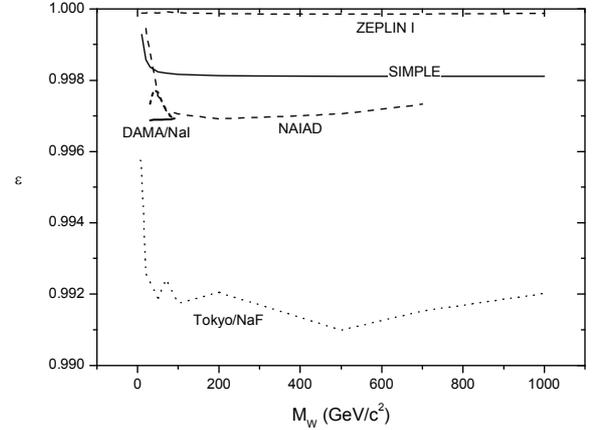

FIG. 12: eccentricities for several experiments of Table I.

The eccentricities of several multi-nuclide spin-dependent experiments with respect to $M_W$ are shown in Fig. 12.

This last parameter is especially important for future perspectives, since a low exposure generally means much room for further improvements. CRESST-I, PICASSO, EDELWEISS, CDMS-II, CRESST-II and DAMA/Xe-2 are not included because single nuclide experiments have by definition $\varepsilon = 1$ due to their infinite $b_M$. Among the quoted experiments, Tokyo/NaF is the most balanced, as a result of the opposite sign of the F and Na $\frac{<S_p>}{<S_n>}$ ratios.

As made clear in Fig. 8, the most restrictive limits on the allowed region of the $a_p$ - $a_n$ plane derive from the intersection of various experiment conics, in particular those with quasi-perpendicular orientations. Note that the odd N experiments are generally quasi-vertically oriented, while in contrast the odd Z experiments are quasi-horizontal. In the odd group approximation, all existing experiments are forced to lie either vertically or horizontally. An intermediate orientation needed, as pointed out in Section IV, to most completely determine the spin-dependent WIMP coupling strengths when positive signals are found, can be achieved by: (i) using odd-odd nuclei, or (ii) combining both odd-Z and odd-N nuclei in the same sensitive material. To date, no current experiment has either of these characteristics.

## VI. CONCLUSIONS

Limits on any spin-dependent WIMP-nucleon interaction at tree level can be reported in a model-independent fashion within one of two equivalent representations: cross section and coupling strength.

In either representation, single isotope experiments like DAMA/Xe-2 leave some WIMP candidates unbounded, due to WIMP-nucleus cross section suppression when the coupling strength ratio approaches a nuclide-dependent "critical" value. Near this value, higher order corrections to the scattering amplitude might become relevant for single nuclide experiment analysis.

Multi-nuclide experiments do not suffer from this complication, because the critical values of different nuclides are generally different, and for $a_{p,n}$ values such that the cross section of one nuclide is quenched to $\sim 0$, there is always another nuclide which maintains its WIMP-sensitivity. For this reason, the protuberance of the allowed region for $\frac{a_p <S_p>}{a_n <S_n>} < 0$ becomes finite.

With this approach, equivalent current limits on the WIMP-nucleon interaction at $M_W$=50 GeV/c² are either $\sigma_p \leq 0.7$ pb; $\sigma_n \leq 0.2$ pb or $|a_p| \leq 0.4$; $|a_n| \leq 0.7$, depending on the choice of representation. These limits become less restrictive for either larger or smaller $M_W$; they are less restrictive than those from the traditional odd group approximation regardless of $M_W$. Surprisingly, experiments traditionally considered spin-independent are seen to severely restrict the spin-dependent phase space.

An experiment's ability to set stringent limits on all tree level theoretical scenarios can be discussed at design time by considering the relative sensitivity D and absolute spin values of available nuclei, or *a posteriori* by examining the actual ε of the experiment. Due to the rarity and instability of odd-odd nuclei, there is little hope to find a nuclide with D~1 optimal for all theoretical scenarios. For a multi-nuclide experiment, the condition D~1 becomes ε ~ 1, which requires a combination of odd Z nuclei and odd (A-Z) nuclei, some with positive spin ratio and others with negative spin ratio. Since up to now there is no experiment with such characteristics, the most theoretically useful information is obtained by intersecting, in either representation, the regions allowed by different experiments. The coupling strength representation makes it clear that this is most efficient when the experiments' ellipses/bands are approximately perpendicular.

When a positive signal is obtained, the regions permitted by a single experiment become shells. A single experiment's positive result can be improved to a curve and not to a point as desirable, which makes intersection of more than two experiments necessary to get restrictive results about the WIMP couplings; two experiments are not enough, because their intersection is four points in the coupling strength representation, and two even in the cross section representation. On the other hand, if there is a single WIMP specie, there must be a region permitted by all experiments with positive signal. If there is no such common permitted region, the existence of a single WIMP specie is excluded.

Of course, there might be two or more WIMP species, such that different experiments are sensitive to different species. This kind of argument has been explored [24,25] in the form of a WIMP doublet with a small separation in mass, so to allow for inelastic scattering of the lighter WIMP to the heavier one. In such a scenario, CDMS-I and DAMA/NaI results for spin-independent WIMPs are not in contradiction [24].

Recently, the results of indirect WIMP searches have begun to yield limits on $\sigma_p$ two orders of magnitude more stringent than those here obtained from direct searches. These derive from model-dependent analyses based on WIMP scattering with solar hydrogen (hence yield a quasi-horizontal band in the coupling strength representation, perpendicular to CDMS-II/ZEPLIN), as well as assumptions regarding the poorly known WIMP annihilation channels; these must be considered with some circumspection, and have not been herein considered. Moreover, focus on only direct detection results allows us to include also the less popular case of Dirac spinor candidates. In fact, a Dirac spinor does not self-annihilate, so it is not excluded by the null solar neutrino observations. On the other hand, a Dirac spinor interacting with ordinary matter in a predominantly spin-dependent fashion is not yet excluded by current direct spin-dependent WIMP searches.

## Acknowledgments

F. Giuliani is supported by grant SFRH/BPD/13995/2003 of the Portuguese Foundation for Science and Technology (FCT). This work has been supported by grant POCTI/FNU/43683 of FCT, co-financed by FEDER.